\documentclass[twocolumn,showpacs,showkeys,preprintnumbers,amsmath,amssymb,superscriptaddress,pre]{revtex4}
\usepackage{graphicx}
\usepackage{dcolumn}
\usepackage{bm}
\usepackage{xcolor}
\renewcommand{\vec}[1]{{\boldsymbol{#1}}}

\begin{document}
\title{QED cascades induced by circularly polarized laser fields}
\author{N.~V.~Elkina}
\affiliation{Ludwig-Maximilians Universit\"{a}t M\"unchen, 80539, Germany}
\author{A.~M.~Fedotov}
\affiliation{National Research Nuclear University MEPhI, Moscow, 115409, Russia}
\author{I.~Yu.~Kostyukov}
\affiliation{Institute of Applied Physics, Russian Academy of Sciences, 603950, Nizhny Novgorod, Russia}
\author{M.~V.~Legkov}
\affiliation{National Research Nuclear University MEPhI, Moscow, 115409, Russia}
\author{N.~B. Narozhny}
\affiliation{National Research Nuclear University MEPhI, Moscow, 115409, Russia}
\author{E.~N. Nerush}
\affiliation{Institute of Applied Physics, Russian Academy of Sciences, 603950, Nizhny Novgorod, Russia}
\author{H.~Ruhl}
\affiliation{Ludwig-Maximilians Universit\"{a}t M\"unchen, 80539, Germany}

\begin{abstract}

The results of Monte-Carlo simulations of electron-positron-photon cascades initiated by slow electrons  in circularly polarized fields of ultra-high strength are presented and discussed. Our results confirm previous qualitative  estimations [A.~M. Fedotov, et al., PRL 105, 080402 (2010)] of the formation of cascades. This sort of cascades has revealed the new property of the restoration of energy and dynamical quantum parameter due to the acceleration of electrons and positrons by the field and may become a dominating feature of laser-matter interactions at ultra-high intensities. Our approach incorporates radiation friction acting on individual electrons and positrons.
\end{abstract}

\maketitle

\section{Introduction}
\label{Sec:Intro}

The dramatic progress in laser technology has enabled a novel area of studies exploring laser-matter interactions at ultra-high intensity \cite{sl_physics}. The intensity level of $2\times 10^{22}\mbox{W/cm}^2$  has recently been achieved \cite{Icurrent} and two projects \cite{HiPeR,ELI} aiming at intensity levels up to $10^{26}\mbox{W/cm}^2$ have been supported and are under way. Furthermore, several original proposals have been suggested e.g. \cite{Mourou,Bulanov,Pukhov}, which reach even higher intensities with almost the present level of technology. One of the key phenomena of laser-matter interactions, that probably dominates at ultra-high intensities of our interest, is  the occurrence of QED cascades \cite{ELI,BellKirk,BellKirk1,Fed}. These cascades (also called avalanches, or showers) are caused by successive events of hard photon emissions and electron-positron pair photoproduction by hard photons. As predicted in \cite{Fed}  based on qualitative estimations, the cascades may arise as soon as the laser field strength exceeds the threshold value of $E_*=\alpha E_S$, where $\alpha=e^2/\hbar c\approx 1/137$ is the fine structure constant and  $E_S=m^2c^3/e\hbar=1.32\times 10^{16}\mbox{V/cm}$ is the characteristic QED field.  Such a field strength corresponds to an intensity  of $\sim 10^{25}\mbox{W/cm}^2$.

Previously  QED cascades have been observed and studied as a part of Extensive Air Showers (EAS) in the context of the passage of ultra-high energy particles, that originate from Cosmic Rays, through the atmosphere \cite{Auger,recent,math}. However, similar processes can be observed in the external electromagnetic field as well. In this case, Bremsstrahlung is replaced by the non-linear Compton scattering and Bethe-Heitler process
is replaced by the non-linear Breit-Wheeler process. The latter processes are well studied both  theoretically \cite{Riess,Ritus,BK,Gold,Baier,Erber} and in laser experiments \cite{SLAC2} and are probably of great importance for astrophysics (see, e.g., \cite{Usov}).

An important novel distinctive feature of the cascades in the ultra-strong laser field, compared to situations ever studied previously, is that the laser field is not only able to be a target for ultra-relativistic electrons and hard photons, but can also accelerate the charged particles to ultra-relativistic energies.

As a result, the cascades can be produced even by initially slow electrons or positrons, if they were somehow injected into the strong field region. Moreover, the mean energy of the particles is no longer decreasing in the course of the cascade development due to its redistribution among the permanently growing number of the created particles, but rather is restoring at the expense of the energy extracted from the laser field. This must lead to a vast increase of the cascade yield, as compared to the cascades in media or in strong magnetic fields. In this case the cascade multiplicity  would be restricted either by the duration of stay of the particles in the focal region of the laser field, or even, under more extreme conditions, by the total energy stored in the laser field. In the latter case the focused laser pulses would be depleted by cascade production.

As it will be explained in more details below, the restoration mechanism works if the particles can be accelerated transversely to the field. It was conjectured \cite{BellKirk,Fed} on the basis of qualitative analysis for the model of a uniformly rotating electric fields, that this may be indeed the case.

In the EAS theory, the 1D approximation is often used because spreading in transverse direction is inessential for ultrarelativistic particles and has no significance for that problem. Besides, the cascade equations can be solved in this case analytically within the ultra-relativistic approximation by means of the Mellin transform \cite{recent,math}. The results of such analytic theory are in good agreement with both experiments \cite{Auger,recent} and direct Monte-Carlo simulations \cite{Messel}. The attempts to treat the cascades in strong magnetic fields on similar grounds are also known \cite{Akhiez}. However, though the $1D$ approximation remains valid, the cascade equations can not be simplified via the Mellin transform unless some further approximation is made. According to Monte-Carlo simulations \cite{bolg}, the resulting analytical approach works much worse here than in the case of 1D approximation for EAS. In our case of cascades arising in a laser field, the structure of the cascade equations (see  Appendix \ref{sec:cascade_eqs}) is the same as for the magnetic field, but it is impossible to incorporate restoration mechanism within the 1D approximation in momentum space. This means that our problem is essentially two- or three dimensional.

In this work we report on the first results of the Monte-Carlo simulations of cascades produced by initially slow electrons in a uniformly rotating homogeneous electric field. Such a field can be obtained practically at the antinodes of a standing electromagnetic wave. The choice of the field model is uniquely specified by the existence of reasonable qualitative estimations for
scaling of the basic cascade characteristics for this particular case \cite{Fed}. Our goal was to prove explicitly the existence of the restoration mechanism and to test the estimations \cite{Fed} by direct numerical simulations.

The paper is organized as follows. In Sec.~\ref{Sec:2}, which can be considered as a technical introduction, we review and collect the known information on the elementary quantum processes: single photon emission by electrons and pair creation by hard photons in  strong fields of arbitrary configuration. Though this information is not completely new, it is of essential importance for our presentation and is spread among the literature on the subject. After that, in Sec.~\ref{Sec:3} we present the reasoning in favor of the energy restoration mechanism for cascades in electromagnetic fields. In Sec.~\ref{Sec:4} we formulate the assumptions of our model, present the details of our Monte-Carlo routine and discuss the results obtained by numerical simulations. These results are compared to the known estimations. Summary and discussion is given in Sec.~\ref{Sec:Discuss}. Finally, in Appendix \ref{sec:cascade_eqs}, we discuss the cascade equations for our problem and explicitly demonstrate that, contrary to the recent doubts \cite{Bul}, the approach we use takes proper account of radiation friction by ultrarelativistic electrons.

\section{Quantum processes with high-energy particles in a strong electromagnetic field}\label{Sec:2}

General properties of radiation of ultrarelativistic particles are well known \cite{Landau2}. Due to the relativistic aberration effect the momenta of the products of any decay of an ultrarelativistic particle are directed  within the narrow angle $\Delta\theta\sim\gamma^{-1}$ with its momentum, where
$\gamma = \sqrt{1 + (\vec p/mc)^2}$. Thus, radiation of a charged ultrarelativistic particle is visible at a point of observation only for a short period of time $\tau$ during which its momentum turns through the angle of the order $\Delta\theta$. The momentum turning angle can be estimated by
$\Delta\theta\sim eF_\perp\tau/mc\gamma$, where $F_\perp$ denotes the characteristic value of the transverse component of the field. Thus, $\tau\sim mc/eF_\perp$. The characteristic frequency $\Omega$ of classical radiation can be most simply estimated in the proper reference frame of the particle, by transition to which the duration  $\tau$ transforms into $\tau'=\tau/\gamma$. Thus, $\Omega'\sim {\tau'}^{-1}\sim eF_\perp\gamma/mc$. In the laboratory frame, due to the Doppler up-shift, we would thus have $\Omega\sim (eF_\perp/mc)\gamma^2$. Such a scaling with $\gamma$ is typical for congeneric problems and arises e.g. in the theory of synchrotron radiation.

The parameter $\chi=\hbar\Omega/(\gamma mc^2)=F_\perp\gamma/E_S$ \cite{Landau4}, being the ratio of the classically estimated mean energy of an emitted photon to the energy of the radiating particle, determines whether the process of radiation is controlled by classical electrodynamics or QED. Namely, if $\chi\ll 1$, then quantum recoil is inessential and radiation is classical, whereas if  $\chi\gtrsim 1$ then it must be quantum. The parameter $\chi$ is Lorentz and gauge invariant and is precisely defined as $\chi=e\hbar/(m^3c^4)\sqrt{-(F_{\mu\nu}p^\nu)^2}$, where $F_{\mu\nu}$ is the strength tensor of the field and $p^\mu$ is the $4$-momentum of a particle. In what follows, we assume that $F\ll E_S$. On the other hand, we assume that the field is of relativistic strength in the sense that the dimensionless field amplitude $a_0=e\sqrt{-A_\mu A^{\mu}}/(mc)\gg 1$, where
$A_\mu$ is the 4-vector of the field potential. The latter means, in particular, that it varies on the scale that exceeds $\tau$ and thus can be considered constant with respect to the decay processes.

Two different theoretical approaches have been developed in order to study photon emission by ultrarelativistic ($\gamma\gg 1$) charged particles in electromagnetic fields of ultrarelativistic ($a_0\gg 1$, $\chi\gtrsim 1$) but still subcritical ($F\ll E_S$) intensities. Nikishov and Ritus (NR) have calculated the appropriate quantum amplitudes in terms of Volkov solutions in a constant \emph{crossed} field ($E^2-H^2=0$ and $\vec{E}\cdot\vec{H}=0$) of arbitrary strength \cite{Ritus,Landau4}. Their results can be applied directly to our problem because, as they have pointed out especially, under the abovementioned conditions \emph{any} field looks locally as constant and crossed, the latter in the sense that both field invariants $(E^2-H^2)/E_S^2$ and $\vec{E}\cdot\vec{H}/E_S^2$ are much less than $\chi^2$. The other approach by Baier and Katkov (BK) \cite{Baier,Landau4} is based on the observation that the motion of a particle in between two acts of photon emission (which corresponds to free lines of Feynman diagrams) can be considered classically if $F\ll E_S$, so that all the relevant quantum corrections are reduced to quantum recoil and, possibly, to field-spin interactions. In a sense, the BK approach is equivalent to the replacement of the aforementioned Volkov solutions by the localized wave packets moving along the classically prescribed trajectories. Both approaches are based essentially on the same approximations and provide the same results for the energy spectra of emitted photons.

The energy distribution of the probability rate for photon emission by ultrarelativistic electrons in an electromagnetic field is given by \cite{Ritus,Baier,Landau4}
\begin{eqnarray}
\frac{dW_{rad}(\varepsilon_\gamma)}{d\varepsilon_\gamma}=-\frac{\alpha m^2c^4}{\hbar\varepsilon_e^2}
\left\{\int\limits_x^\infty {\rm Ai}\,(\xi)\,d\xi\right.\nonumber\\\left.
+\bigg(\frac2{x}+\chi_\gamma\sqrt{x}\bigg)\,{\rm Ai}'(x)\vphantom{\int\limits_x^\infty}\right\},
\label{phot_emiss}
\end{eqnarray}
where
$x=(\chi_\gamma/\chi_e\chi_e')^{2/3}$, ${\rm Ai}(x)=(1/\pi)\int_0^\infty \cos(\xi^3/3+\xi x)\,d\xi$
is the Airy function, $\varepsilon_\gamma$ and $\varepsilon_e$ are the energies of the emitted photon and the initial electron, respectively. $\chi_e$, $\chi_e'=\chi_e-\chi_\gamma$ and $\chi_\gamma$ $(0<\chi_\gamma<\chi_e)$ are the dimensionless quantum parameters for the electron before and after emission, and for the emitted photon, respectively. In terms of the field strengths, this parameter is represented as
\begin{equation}\label{chi}
\chi=\frac{e\hbar}{m^3c^4}\sqrt{\left(\frac{\varepsilon\vec{E}}{c}
+\vec{p}\times\vec{H}\right)^2-(\vec{p}\cdot\vec{E})^2},
\end{equation}
where $\varepsilon$ and $\vec{p}$ are the energy and the momentum of the corresponding particle.

Note that expression (\ref{phot_emiss}) suffers from the infrared singularity at $\varepsilon_\gamma\to 0$. However, in our constant field approximation this singularity  $dW_{rad}(\varepsilon_\gamma)/d\varepsilon_\gamma=O(\varepsilon_\gamma^{-2/3})$
is weaker than the usual $O(\varepsilon_\gamma^{-1})$ scaling of the infrared behavior of perturbative QED \cite{infrared,Landau4}. In particular, the total radiation probability rate is infrared convergent in our approximation. The infrared sector, however, is not important for the parameters considered in our paper, because most of the emitted radiation are found to have much larger frequencies than the frequency of the driving field.

The energy distribution of the probability rate for direct pair creation by hard photons ($\varepsilon_\gamma\gg mc^2$) is given by \cite{Ritus,Baier,Landau4}
\begin{eqnarray}
\frac{dW_{cr}(\varepsilon_e)}{d\varepsilon_e}=\frac{\alpha m^2c^4}{\hbar\varepsilon_\gamma^2}
\left\{\int\limits_x^\infty {\rm Ai}\,(\xi)\,d\xi\right.\nonumber\\\left.
+\bigg(\frac2{x}-\chi_\gamma\sqrt{x}\bigg)\,{\rm Ai}'(x)\vphantom{\int\limits_x^\infty}\right\},
\label{pair_cr}
\end{eqnarray}
where the indices ``$\gamma$'' and ``$e$'' this time refer to the initial photon and to the created electron, respectively. For the created positron, we have $\chi_e'=\chi_\gamma-\chi_e$ $(0<\chi_\gamma<\chi_e)$. Formula (\ref{pair_cr}) is completely symmetric with respect to electron and positron remaining unchanged under the replacement $\chi_e \leftrightarrow \chi_e'$. Similarity between formulas (\ref{phot_emiss}) and (\ref{pair_cr}) is explained by the fact that these two processes are related by the cross-symmetry \cite{Landau4}.

\begin{figure}
\includegraphics[width=6.5cm]{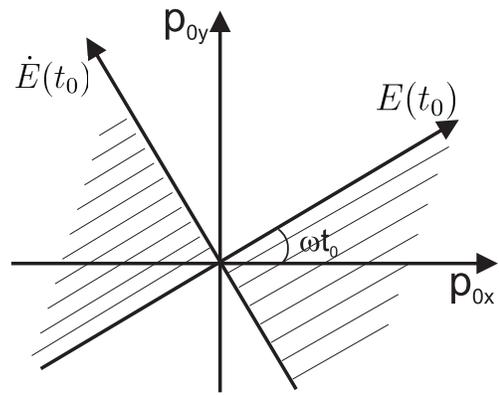}
\caption{\label{accreg} The sign of $\dot{\chi}_e(t)$ along the particle trajectory at $t=t_0$ in different zones of the $p_{0x}p_{0y}$ plane. The shaded zones correspond to acceleration (increase of $\chi_e$) of positrons and deceleration (decrease of $\chi_e$) of electrons. The non-shaded zones correspond to the vice-versa situation.}
\end{figure}

The total probability rates for both processes
\begin{equation}\label{totals}
W_{rad}=\int\limits_0^{\varepsilon_e}
\frac{dW_{rad}(\varepsilon_\gamma)}{d\varepsilon_\gamma}\,d\varepsilon_\gamma,\quad
W_{cr}=\int\limits_0^{\varepsilon_\gamma} \frac{dW_{rad}(\varepsilon_e)}{d\varepsilon_e}\,d\varepsilon_e,
\end{equation}
cannot be written in terms of known special functions and should be obtained by numerical integrations. However, they allow simple asymptotic expressions in the limits of small and large $\chi_e$, $\chi_\gamma$, respectively. Namely, we have
\begin{subequations}\label{Wrad}
\begin{equation}
W_{rad}\approx
1.44\frac{\alpha m^2c^4}{\hbar\varepsilon_e}\chi_e,\quad\chi_e\ll 1,\label{Wrad_small_chi}
\end{equation}
\begin{equation}
W_{rad}\approx 1.46\frac{\alpha m^2c^4}{\hbar\varepsilon_e}\chi_e^{2/3},\quad\chi_e\gg 1,
\label{Wrad_large_chi}
\end{equation}
\end{subequations}
and
\begin{subequations}\label{Wcr}
\begin{equation}\label{Wcr_small_chi}
W_{cr}\approx
0.23\frac{\alpha m^2c^4}{\hbar\varepsilon_\gamma}\chi_\gamma e^{-8/3\chi_\gamma},\quad\chi_\gamma\ll 1,
\end{equation}
\begin{equation}\label{Wcr_large_chi}
W_{cr}\approx 0.38\frac{\alpha
m^2c^4}{\hbar\varepsilon_\gamma}\chi_\gamma^{2/3},\quad\chi_\gamma\gg 1.
\end{equation}
\end{subequations}
Eqs.~(\ref{Wrad_small_chi}), (\ref{Wcr_small_chi}) in these formulas describe the quasiclassical regime. For small values of the quantum parameter $\chi_\gamma$, the probability rate for pair photoproduction $W_{cr}$ is suppressed exponentially, in accordance with the essentially quantum nature of this process. At the same time, $W_{rad}$ remains $O(\hbar^{-1})$, thus providing a finite classical limit for the mean radiated intensity $I_{rad}$.
As for the limit of large $\chi$, both rates (\ref{Wrad_large_chi}) and (\ref{Wcr_large_chi}) differ only by a numerical factor of the order of unity.

Given the energy and the momentum of the electron before emission, Eq.~(\ref{phot_emiss}) determines the probability distribution for the
energy $\varepsilon_\gamma$ of the emitted photon. Under our assumption $\gamma_e\gg 1$, the momentum of this photon is given by $\vec{p}_\gamma=(\varepsilon_\gamma/p_e)\vec{p}_e$. The energy and the momentum of the electron after emission should be determined from the conservation laws. In the electromagnetic background, they are of the form $p_e^\mu+q^\mu={p_e'}^\mu+p_\gamma^\mu$, where $q^\mu$ is the four-momentum extracted from the field. The exact value of this $q^\mu$ essentially depends  on the global structure of the field. This is because the whole spacetime contributes to the integrals in the QED matrix element that yield delta-functions expressing the conservation laws. For example, in a crossed constant field with the Poynting vector directed along the $z$-axis the conserved quantities are $\varepsilon/c-p_z$, $p_x$ and $p_y$. In a constant electric field, the canonical momentum is conserved. In a constant magnetic field, directed along the $z$ axis and with symmetric gauge, the conserved quantities are the number of the Landau level, the angular momentum, and $p_z$. Nevertheless, for ultrarelativistic particles there is actually no difference between these  possibilities, and either of them can be adopted with the accuracy of our approximation. The reason is that $q\lesssim eF\tau\lesssim mc\ll p_e,p_e',p_\gamma$. In particular, we can assume $\vec{p}_e'=\vec{p}_e-\vec{p}_\gamma$. The same argument can be applied to pair creation by hard photons as well.

In addition to one-photon emission and direct pair photoproduction reviewed above, there exist more complicated higher-order processes, such as e.g. the two-photon emission $e^-\to e^-\gamma\gamma$ or the trident process $e^-\to e^-e^-e^+$. Their specific feature is that the intermediate particle is off the mass shell, i.e. is virtual. However, in strong field situations of our interest the two-step processes dominate \cite{SLAC2,BellKirk,BellKirk1}. For this reason, we do not consider higher-order processes in the sequel.

\section{Basic estimations for cascade production in a rotating electric field}
\label{Sec:3}

\begin{figure}
\includegraphics[width=8.6cm]{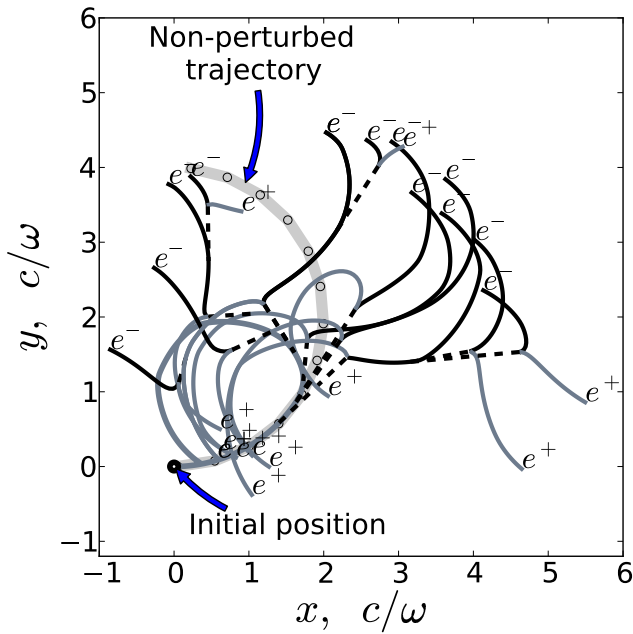}
\caption{\label{cascade_devel} Spatial picture of the formation of the cascade initiated by a positron in the homogeneous uniformly rotating electric field (obtained by a Monte-Carlo simulation with $a_0=2\times 10^3$ and $\hbar\omega=1\mbox{eV}$). Legend: Trajectories of electrons and positrons are shown as black and gray curves, respectively. The hard photons which have created pairs during the simulation time are shown as the dashed lines. The trajectory of the primary positron ignoring any QED processes is plotted as the thick light gray curve.}
\end{figure}

Since there is no difference whether an electron or a positron  initiates a cascade, we  assume in this section  that our cascade is initiated by a positron ($e>0$).

Consider a positron in a homogeneous, uniformly rotating electric field
\begin{equation}\label{E(t)}
\vec{E}(t)=\{E_0\cos{\omega t},E_0\sin{\omega t}\}.
\end{equation}
The equation of motion
\begin{equation}\label{Newton}
\dot{\vec{p}}(t)=e\vec{E}(t),
\end{equation}
with the initial condition $\vec{p}(t_0)=\vec{p}_0$, can be easily solved:
\begin{eqnarray}
p_x(t)=p_{0x}+mca_0(\sin{\omega t}-\sin{\omega t_0}),\nonumber\\
p_y(t)=p_{y0}-mca_0(\cos{\omega t}-\cos{\omega t_0}).\label{Newton_sol}
\end{eqnarray}
Here, $a_0=eE_0/m\omega c$ is the dimensionless field amplitude.

Let us assume first that the positron is at rest ($\vec{p}_0=0$) initially ($t_0=0$). Equations~(\ref{E(t)}) and~(\ref{Newton_sol}) show that the energy and the quantum parameter $\chi$ of the positron for this case depend on time as
\begin{subequations}
\begin{equation}\label{eps(t)}
\varepsilon_e(t)=mc^2\sqrt{1+4 a_0^2\sin^2{\frac{\omega t}{2}}},
\end{equation}
\begin{equation}\label{chi(t)}
\chi_e(t)=\frac{e\hbar E_0}{m^2c^3}\sqrt{1+4 a_0^2\sin^4{\frac{\omega t}{2}}}.
\end{equation}
\end{subequations}
Both quantities are increasing initially. They are oscillating with the period $2\pi/\omega$ of the rotation of the field. The amplitudes of these oscillations, $\varepsilon_m\approx 2mc^2a_0$, $\chi_m\approx 2a_0(E_0/E_S)=2(\hbar\omega/mc^2) a_0^2$ are proportional to $a_0$ and $a_0^2$, respectively, and are quite large under our basic assumptions. For example, $\chi_m$ approaches unity already at $a_0\sim a_{0c}=500$ for an optical rotation frequency of $\hbar\omega=1\mbox{eV}$. This corresponds to the field strength $E_0\sim 10^{-3}E_S\sim 10^{13}\mbox{V/cm}$ and the intensity $10^{24}\mbox{W/cm}^2$. Since $\chi_m\sim 1$ under such conditions, the positron, according to the preceding section, is able to emit a hard photon with $\chi_\gamma\sim \chi_e\sim 1$, which, in turn, can create an electron-positron pair. However, at such intensities a new generation of pairs is typically produced on the time scale $\pi/\omega$, and the whole pair generation process may be rather sensitive to peculiarities of the field model. As we discuss below, stable cascade formation is expected at higher intensity levels.

The formulas (\ref{eps(t)}) and (\ref{chi(t)}) become especially simple for stronger fields $a_0\gg a_{0c}$, because in this case the value $\chi_e\sim 1$ is being reached within just a small fraction $t_{acc}$ of the rotation period. Namely, we have
\begin{subequations}
\begin{equation}\label{eps(t)1}
\varepsilon_e(t)\approx eE_0 ct,\quad \frac1{\omega a_0}\ll t\ll\frac1{\omega},
\end{equation}
\begin{equation}\label{chi(t)1}
\chi_e(t)\approx \frac12\left(\frac{E_0}{E_S}\right)^2\frac{mc^2\omega}{\hbar}t^2,\quad
\frac1{\omega \sqrt{a_0}}\ll t\ll\frac1{\omega}.
\end{equation}
\end{subequations}
Eq.~(\ref{eps(t)1}) is easy to understand, because initially the positron is accelerating almost along the field. In order to understand Eq.~(\ref{chi(t)1}) better, let us note that in the case $\vec{p}(0)=0$, according to Eqs.~(\ref{E(t)}) and (\ref{Newton_sol}), the momentum of the positron constitutes exactly the angle $\omega t/2$ with both the instant direction of the field and the $x$-axis. Intuitively, this is because due to its inertia the particle does not follow the rotation of the field precisely. As a consequence, the transverse component of the field with respect to momentum of the particle increases as $E_\perp=E_0\sin(\omega t/2)\approx E_0\omega t/2$. Since $\chi_e(t)\approx E_\perp\gamma/E_S$, we arrive immediately at Eq.~(\ref{chi(t)1}). Qualitatively, the same growth of the energy and the parameter $\chi$ with time has been observed for generic field configurations \cite{Fed}.

\begin{figure}
\includegraphics[width=8.0cm]{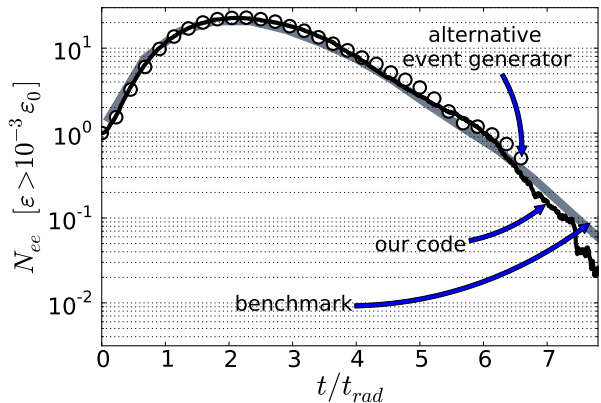}
\caption{\label{compare_ang} Comparison of the cascade profile obtained with our code [black thin line], with a code applying an alternative event generator [circles] and from previous independent simulations [thick gray line, see Fig.~5 in \cite{bolg}]. Depicted is the number of pairs with energy $\varepsilon>10^{-3}\varepsilon_0$ versus the elapsed time. The simulation parameters are $\varepsilon_0=100\mbox{GeV}$ and $E_0/E_S=0.2$.}
\end{figure}

As it follows from Eq.~(\ref{chi(t)1}), the quantum parameter $\chi_e$  becomes of the order of unity over the period of time $t_{acc}$,
\begin{equation}\label{t_acc}
t_{acc}\sim \frac{\hbar}{\alpha mc^2\mu}\sqrt{\frac{mc^2}{\hbar\omega}}.
\end{equation}
Here we have introduced a new dimensionless field intensity parameter $\mu=E/E_*$, $E_*=\alpha E_S\approx E_S/137$, which is appropriate for the cascade problem \cite{Fed}. The parameter $\mu$ is related to the commonly accepted parameter $a_0$ by $\mu=(\hbar\omega/\alpha mc^2) a_0$. According to Ref.~\cite{Fed}, the cascades can be caused by initially slow particles if $\mu\gtrsim 1$.

In the course of hard photon emission, the value of the quantum parameter $\chi_e$ is shared between the positron and the emitted photon \footnote{This conservation low is satisfied exactly in a globally constant crossed field, see the remark in the Sec.~\ref{Sec:2}.}, $\chi_e\approx \chi_\gamma+\chi_e'$. If $\chi_e\gtrsim 1$ then both $\chi_\gamma$ and $\chi_e'$ are less than $\chi_e$ but are of comparable value $\chi_e'\sim\chi_\gamma\lesssim\chi_e$. Although propagation of the resulting hard photon is not affected by the field, nevertheless its $\chi_\gamma$ continues to increase after emission just due to rotation of the field.

In order to understand better what must happen after the first hard photon emission, let us come back to Eq.~(\ref{Newton_sol}) and consider the general initial condition. It is easy to see that in the case $\vec{H}=0$ the sign of the derivative of the quantity (\ref{chi}) is determined completely by the expression $-e(\vec{p}\cdot \vec{E})(\vec{p}\cdot \dot{\vec{E}})$. The zones in the $\vec{p}_0$-plane in Fig.~\ref{accreg}, where $\dot\chi_e>0$ is valid for positrons at time $t=t_0$ are shaded. In the shaded areas $\chi_e$ can be expected to increase for some time. The time $t_0$ can be identified with the creation time of a new pair.

Since the momentum of a primary positron is confined to
the shaded region to the right, and the new secondary particles are created with momenta parallel to the momentum of parental particle in our approximation, we see that momenta of the secondary particles also lie in the shaded sector. Thus, the newly created positrons are accelerating with $\dot\chi(t)>0$, while the newly created electrons are initially decelerating ($\dot\chi<0$). However, they are quickly turned back by the field and also get accelerated. The same will be true for successive pair creation processes as well.

In spite of the complexity of the picture of the cascade development (see Fig.~\ref{cascade_devel}), some general estimations for it can nevertheless be obtained \cite{Fed} in the high-field limit $\mu\gg 1$. The idea is that, due to similarity of the rates (\ref{Wrad_large_chi}) and (\ref{Wcr_large_chi}), and also because the variation of the angles between the momenta of all particles (positrons, electrons, and photons) and the field is determined by the same temporal scale $\omega^{-1}$, there is actually no need to distinguish between all three sorts of particles. So, an order of magnitude estimation can be provided within the model of a simple doubling chain process.

\begin{figure*}
\includegraphics[width=8.0cm]{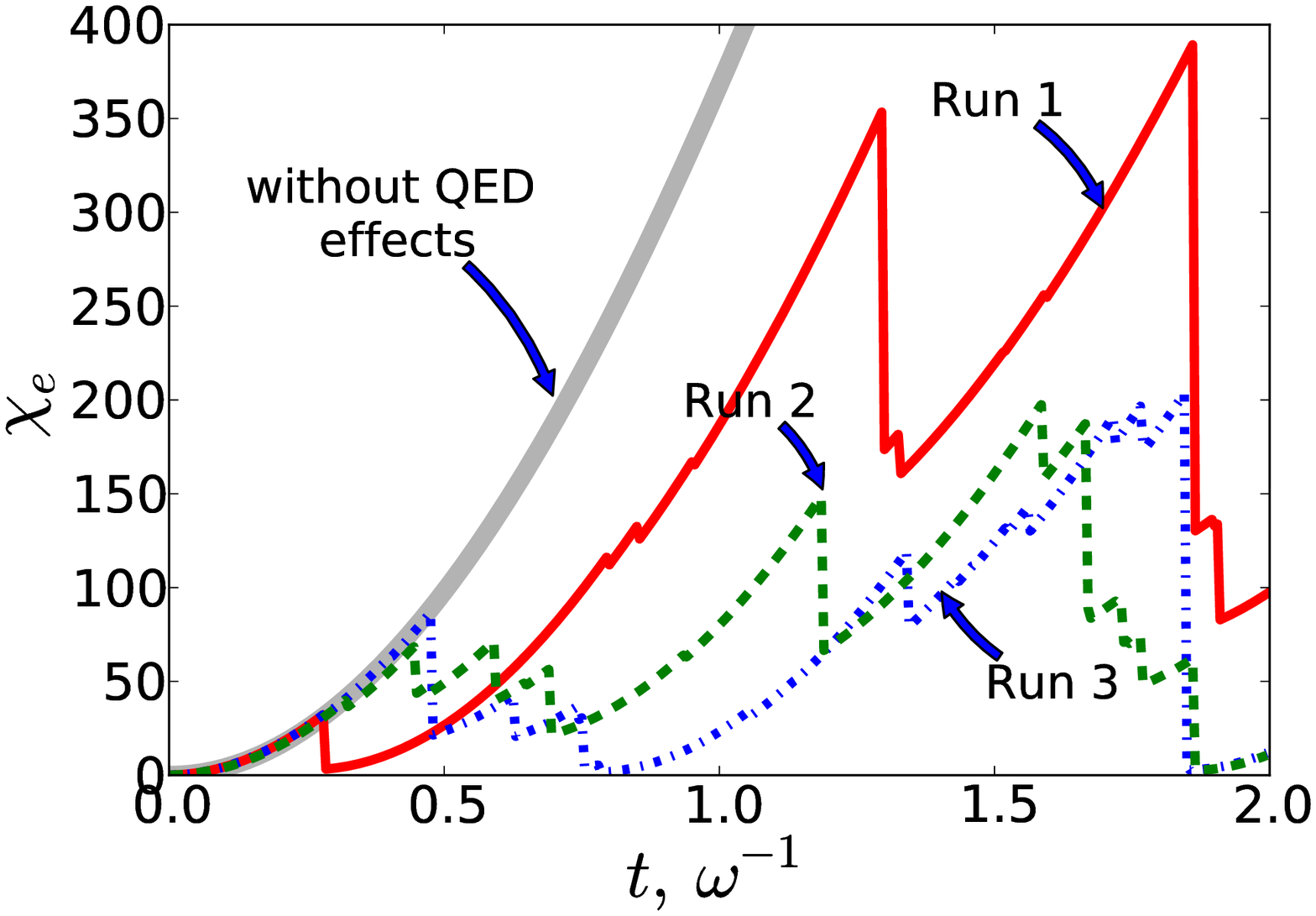}
\includegraphics[width=8.0cm]{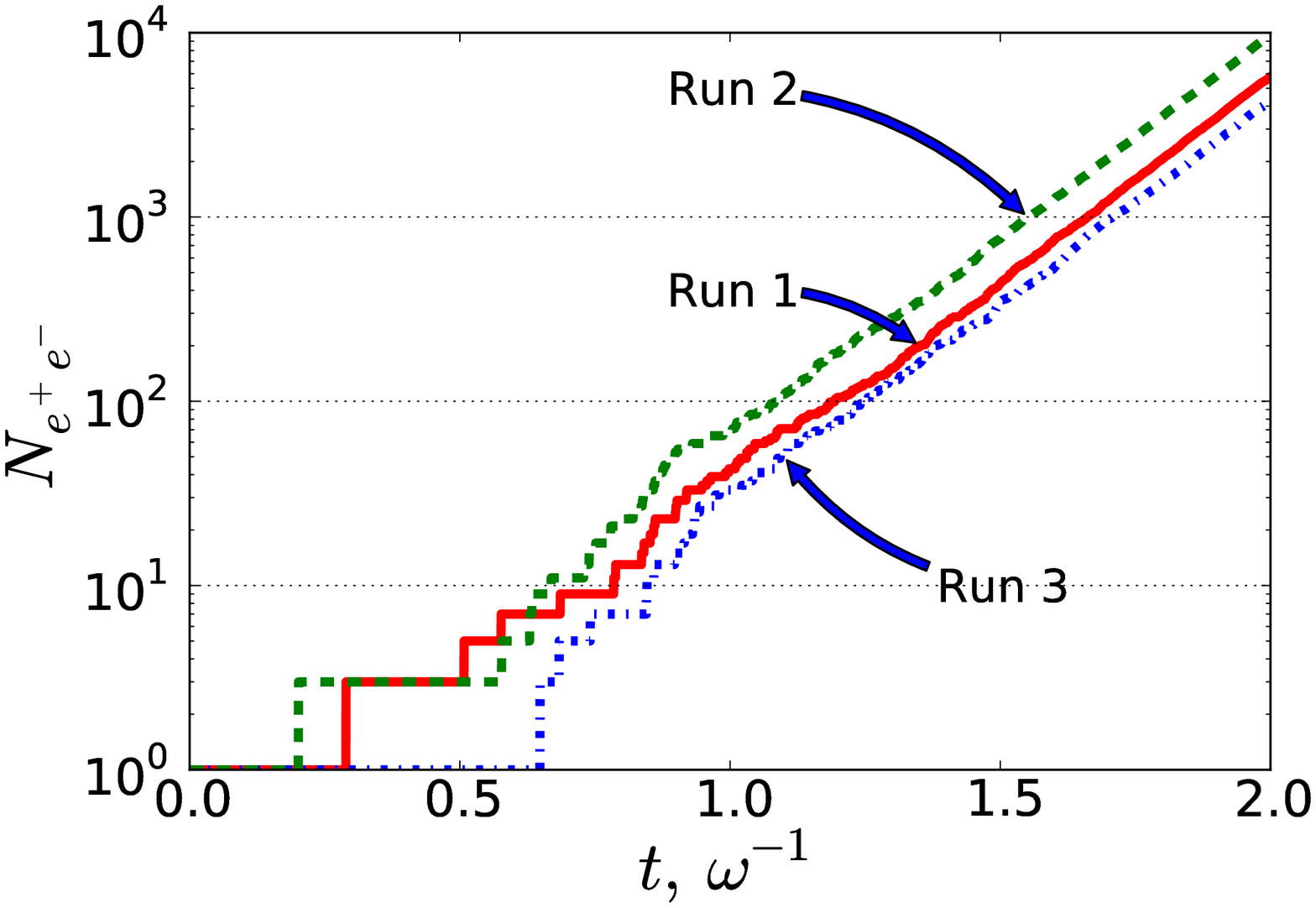}
\caption{\label{chi_N(t)}
Left plot: Temporal evolution of the quantum dynamical parameter $\chi_e$ of the primary electron for three independent Monte-Carlo simulations. The thick gray curve corresponds to the analytical solution Eq.~(\ref{chi(t)}) for $\chi_e(t)$ in the absence of any QED processes. The three other curves (Run1, Run2, and Run 3) are the results of the three independent Monte-Carlo simulations with parameters $a_0=2\times 10^4$ and $\hbar\omega= 1\, \mbox{eV}$. Right plot: The total number of electrons and positrons $N_{e^+e^-}$ vs. time for the same independent simulations.}
\end{figure*}

Let us denote by $t_e$ the typical lifetime for electrons and positrons with respect to hard photon emission. The same quantity up to an order of magnitude defines the lifetime of photons with respect to pair creation. The lifetime $t_e$, together with the typical energy and the value of quantum parameters of the particles, as well as with the angle between their momenta and the field, can be estimated as
\cite{Fed}
\begin{subequations}\label{estimations}
\begin{equation}\label{t_em_mu}
t_e\sim \frac{\hbar}{\alpha mc^2\mu^{1/4}}\sqrt{\frac{mc^2}{\hbar\omega}},
\end{equation}
\begin{equation}\label{energies_chi_mu}
\varepsilon\sim mc^2\mu^{3/4}\sqrt{\frac{mc^2}{\hbar\omega}},\quad \chi\sim\mu^{3/2},
\end{equation}
\begin{equation}\label{theta_mu}
\theta\sim \omega t_e\sim\frac1{\alpha \mu^{1/4}}\sqrt{\frac{\hbar\omega}{mc^2}}.
\end{equation}
\end{subequations}
Under the condition $\mu\gg 1$, as is assumed here, we have $\theta\ll 1$ and $\chi\gg 1$. The latter inequality approves the choice of the asymptotic expressions (\ref{Wrad_large_chi}) and (\ref{Wcr_large_chi}). In addition, we have the following hierarchy of the time scales $t_{acc}\ll t_e$, which assures that exactly hard photons with $\chi_\gamma\gtrsim 1$ are typically emitted.

Within the framework of the doubling chain process model, the number of pairs (multiplicity of the cascade) must grow exponentially,
\begin{equation}\label{N(t)}
N(t)\sim e^{\Gamma t},\quad \Gamma\sim \frac1{t_e}\sim \alpha\mu^{1/4}
\sqrt{\frac{mc^2\omega}{\hbar}}.
\end{equation}
In the next section, we are checking the estimations (\ref{estimations}) and (\ref{N(t)}) by direct Monte-Carlo simulations.

\section{Description of Monte-Carlo approach and numerical results}
\label{Sec:4}

In our simulations we are using a Monte-Carlo approach for the integration of the cascade equations [see the Eqs.~(\ref{casc1}), (\ref{casc2})]. We trace the motion of the electrons and positrons in between the photon emissions classically, whereas for hard photons we exploit the ray tracing approximation in between their emission and conversion into pairs. Even though there exists the exact analytical solution (\ref{Newton_sol}) for equations of motion (\ref{Newton}) for positrons and electrons, we are integrating Eq.~(\ref{Newton}) numerically for each of the particles. This is done in order to incorporate the probabilistic events of photon emission and pair creation in the routines as described below, as well as for the purpose of future generalization to more realistic field configurations.

Our numerical algorithm works as follows. At each time step $t_i<t<t_i+\Delta t$ we are calculating  the momenta of all the particles created at preceding time steps by $\vec{p}_{i+1}=\vec{p}_i+q_i \vec{E}_{i+1/2}\Delta t$, where $\vec{E}_{i+1/2}=\vec{E}(t_i+\Delta t/2)$ and $q_i=+e,-e,0$ for positrons, electrons and photons, respectively. The event generator determines which of the electrons or positrons is going to emit a photon at this time step and whether any of the present photons is going to produce a pair.

\begin{figure*}
\includegraphics[width=8.0cm]{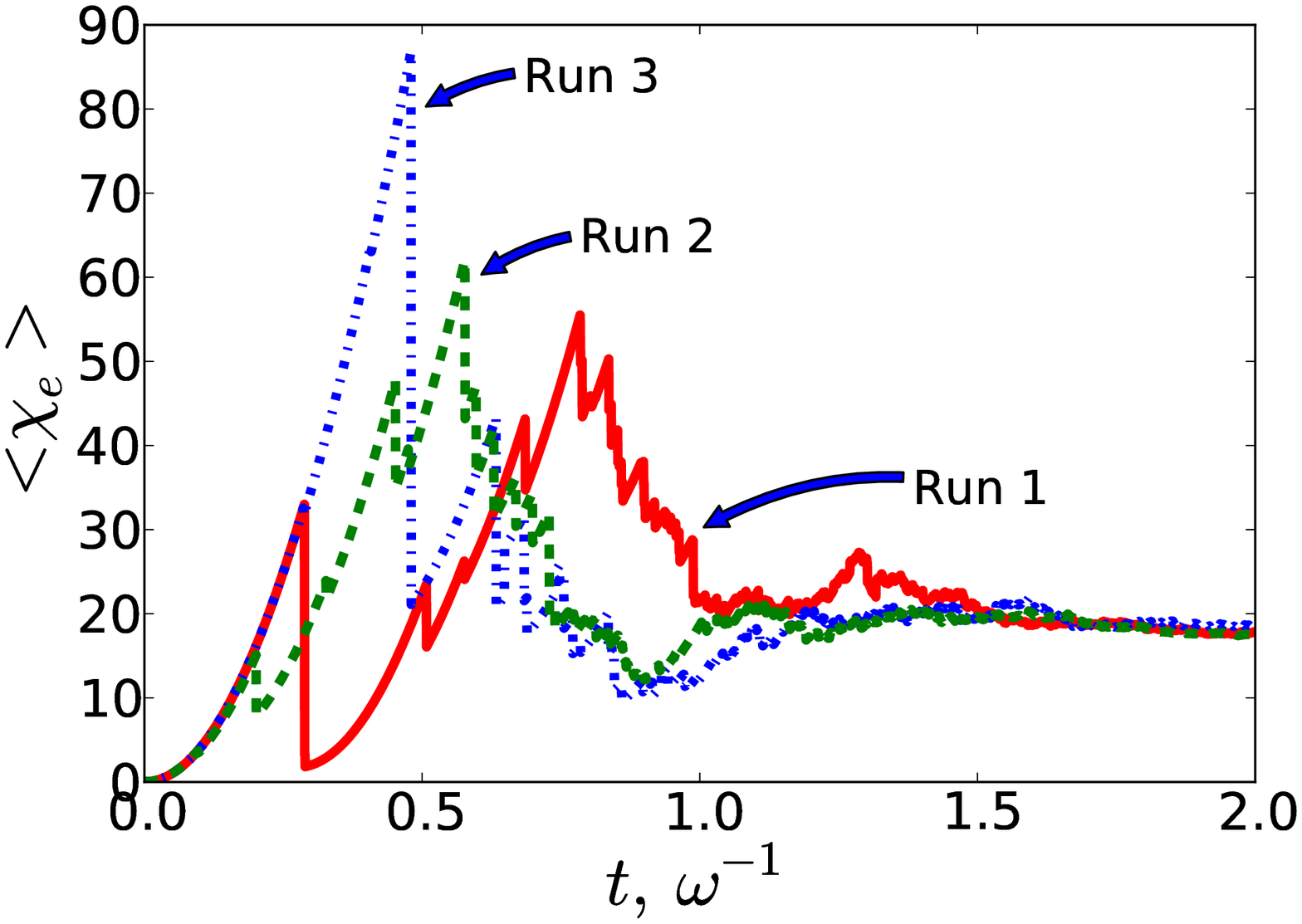}
\includegraphics[width=8.0cm]{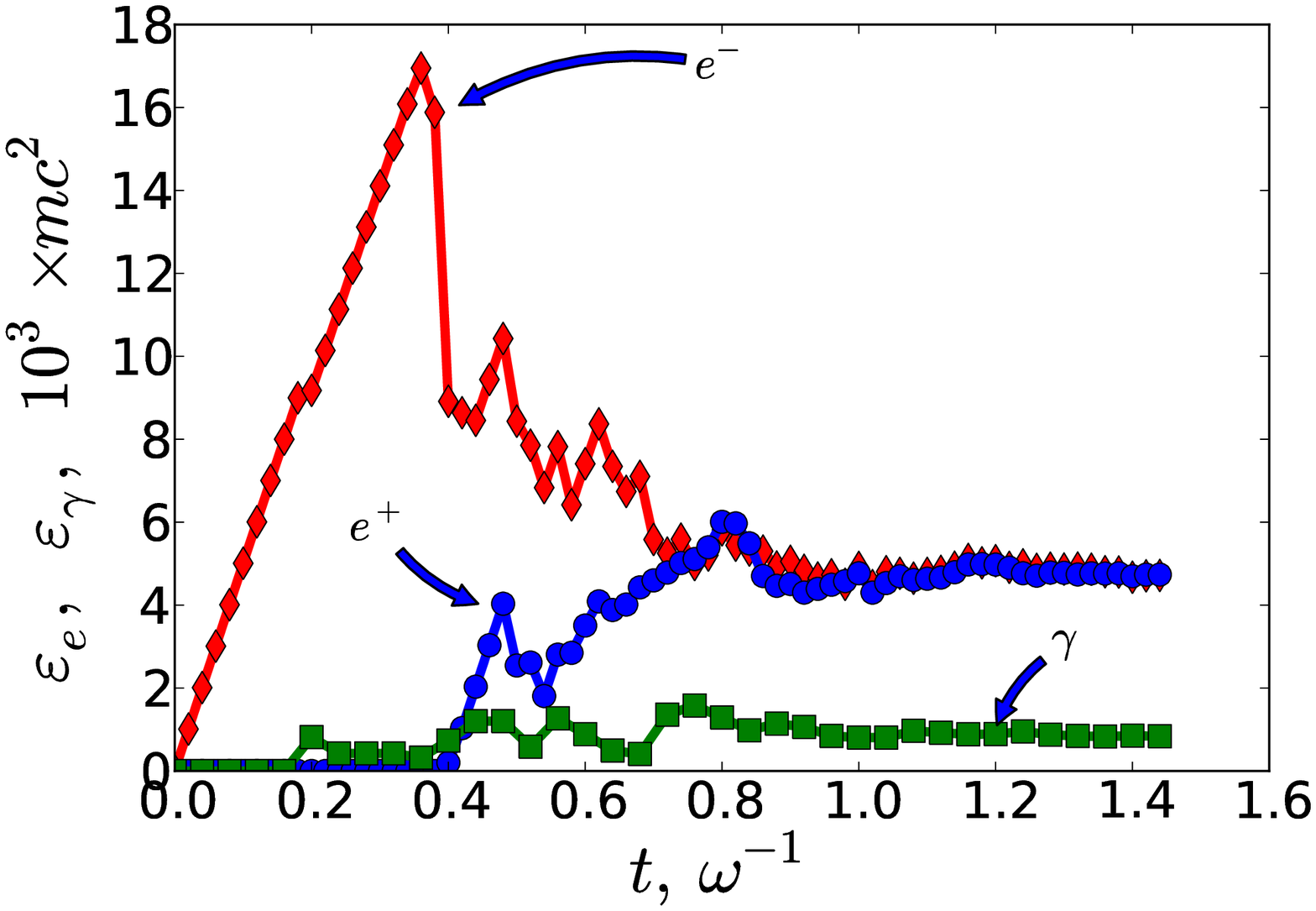}
\caption{\label{av_chi_en}
Left plot: The dynamical quantum parameter $\langle\chi_e\rangle$ for the electrons averaged over the cascade vs. time for the same simulations as in Fig.~\ref{chi_N(t)}. Right plot: Evolution of the mean energy of the electrons, positrons, and photons averaged over the cascade in a typical simulation run ($a_0=5\times 10^4$ and $\hbar\omega=1\mbox{eV}$).}
\end{figure*}

Let us explain our event generator for photon emission in more details (see also Refs.~\cite{Messel,sokolov}). Starting from $\vec{p}_i$ and $\vec{E}_i=\vec{E}(t_i)$, we attach  the value $\chi_i$ at time $t_i$ using Eq.~(\ref{chi}) to each electron and positron and compute the total probability rate $W_{rad}$ (see Eq.~(\ref{totals})). In order to isolate the infrared singularity, we set the lower limit of integration to $\varepsilon_{min}$. For each electron and positron, we assume that it emits a photon between $t_i$ and $t_{i+1}$ if $r<W_{rad}\Delta t$, where $r$ ($0<r<1$) is a uniformly distributed random number. If the above inequality is fulfilled, then the energy $\varepsilon_\gamma$ of the emitted photon is obtained as the root of the sampling equation
\begin{equation}\label{en_phot}
\frac1{W_{rad}}\int\limits_{\varepsilon_{min}}^{\varepsilon_\gamma}
\frac{dW_{rad}(\varepsilon_\gamma)}{d\varepsilon_\gamma}\,d\varepsilon_\gamma=r',
\end{equation}
where $r'$ ($0<r'<1$) is an independent random number. The time step $\Delta t$, which remains fixed in the course of computation, must be chosen such that $\Delta t\ll W_{rad}^{-1},W_{cr}^{-1}$ holds. The direction of propagation of the newly emitted photon is parallel to the momentum $\vec{p}_i$ of the parental electron or positron, whose momentum after emission we find from the conservation law as discussed in Sec.~\ref{Sec:2}. For pair creation, the event generator works similarly, apart from the fact that there is no need for the regularization parameter $\varepsilon_{min}$.

Within the constant crossed field approximation applied here we assume that $\varepsilon_\gamma\gg mc^2$. However, the photons with energies $\varepsilon_\gamma\lesssim mc^2$ are not able to create pairs in a subcritical field, for which $\chi_\gamma\ll 1$ holds. Therefore, we can completely  neglect emission of soft photons in our problem. Based on this reasoning, we currently set the lower integration limit $\varepsilon_{min}$ to $mc^2$ for both $W_{rad}$ and the sampling Eq.~(\ref{en_phot}).

As a benchmark for our code we have simulated the development of a cascade initiated by a high-energy ($\varepsilon_0=2\times 10^5mc^2$) initial electron in a constant homogeneous transverse field with $E_0=0.2E_S$. Our results are averaged over $10^3$ simulation runs. In this particular simulation, the curvature of trajectories of electrons and positrons has been neglected, so that the results of our simulations can be directly compared with previous simulations of cascades produced by high-energy electrons in a magnetic field \cite{bolg}. Comparison of cascade profiles obtained in both simulations is given in Fig.~\ref{compare_ang} by the solid red and dashed green lines, respectively. The figure represents the number of pairs with an energy exceeding $0.1\%$ of the energy of the primary electron versus the elapsed time. In our notation the reference characteristic radiation time $t_{rad}$ as adopted in Ref.~\cite{bolg} in our notation is $t_{rad}=3.85\times (\gamma_{in}/\alpha\chi_{in}^{2/3})\times(\hbar/mc^2)=5.64\times W_{rad,in}^{-1}$, where the subscript ``in'' refers to the initial data for primary electron. We see that our results are in reasonable agreement with the paper \cite{bolg}.

We have also implemented and tested a different event generator, which provides significant speed up due to the absence of numerical integrations. The idea is to exploit some explicit algebraic fits for the energy spectrum (\ref{phot_emiss}), and to exchange the order of testing the occurrence of photon emission and of sampling its energy. In this alternative version of the algorithm, within each time step one first samples the possible energy of an emitted photon just as an uniformly distributed random quantity, $\varepsilon_\gamma=\varepsilon_e r'$ in the above notation. After that, photon emission is assumed to take place if $r<[dW_{rad}(\varepsilon_\gamma)/d\varepsilon_\gamma]\varepsilon_e\Delta t$. In this case, the time step must satisfy the condition $\Delta t<[\varepsilon_e dW_{rad}(\varepsilon_\gamma^*)/d\varepsilon_\gamma]^{-1}$ for all appearing electrons and positrons, where $\varepsilon_\gamma^*$ is the photon energy that corresponds to the maximum of the emission spectrum (\ref{phot_emiss}). The same scheme can be applied to the simulation of pair photoproduction as well. Note that in this case there is no need to introduce the energy cutoff $\varepsilon_{min}$, although this may serve as a useful trick if one wants to restrict the number of soft photons that are traced by the code. The test of the modified event generator is included by a dashed blue line in Fig.~\ref{compare_ang}. This test demonstrates that both versions of the event generator are in fact equivalent.

\begin{figure}
\includegraphics[width=8.0cm]{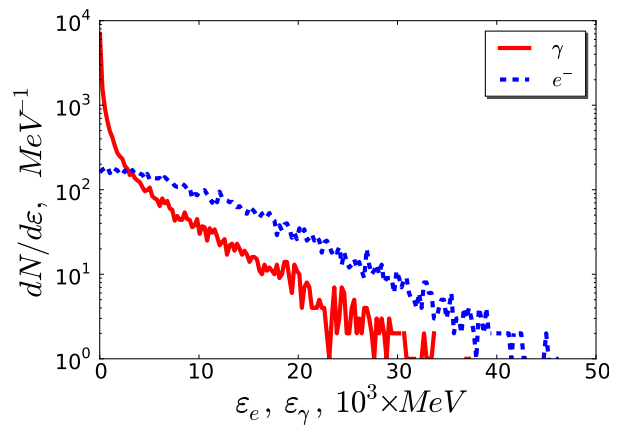}
\caption{\label{spectr}
The energy spectra for different components of the cascade at $t = 1.2\times\omega^{-1}$ for $a_0=5\times 10^4$ and $\hbar\omega = 1\,\mbox{eV}$.}
\end{figure}

\begin{figure*}
\includegraphics[width=8.0cm]{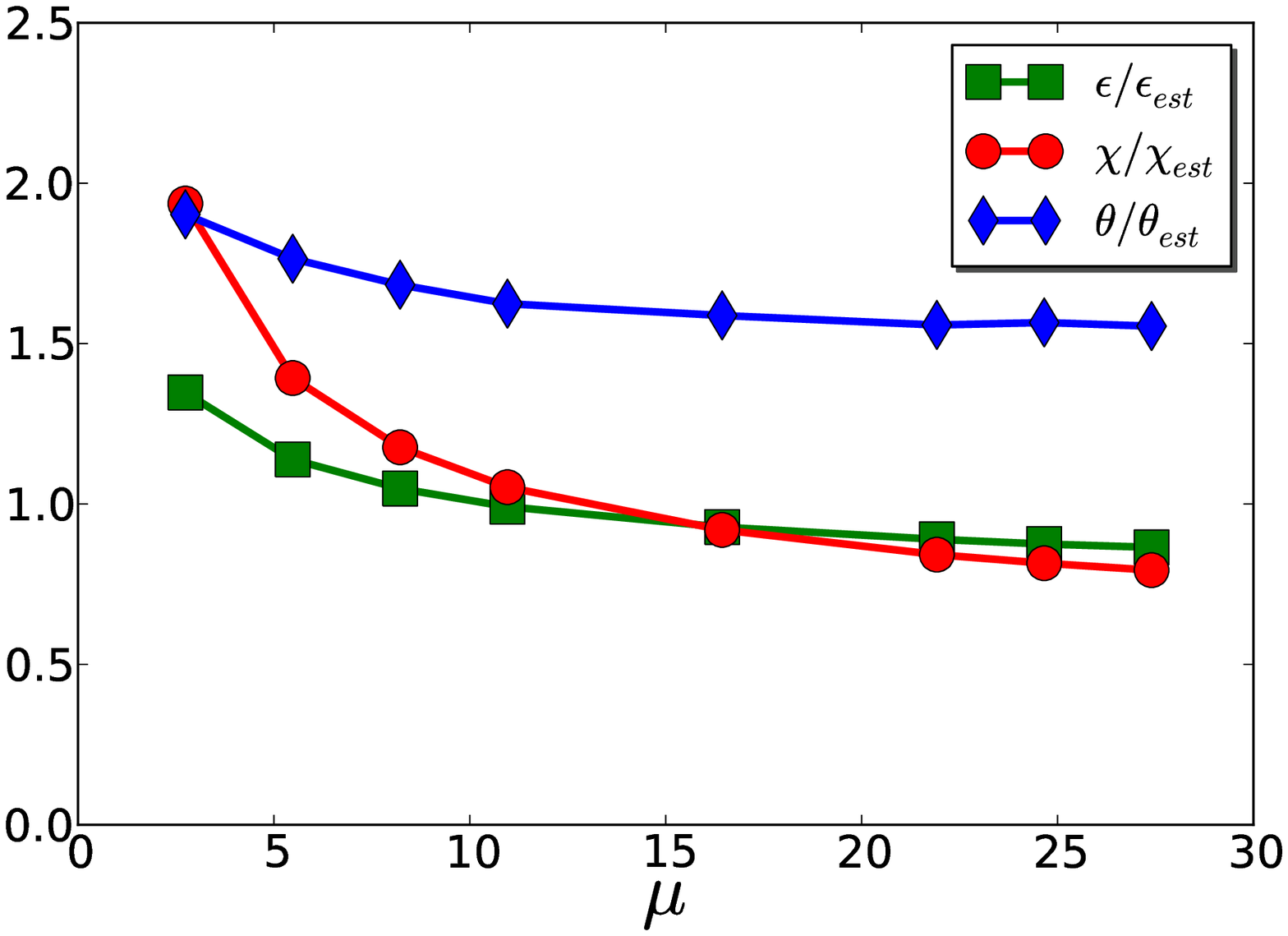}
\includegraphics[width=8.0cm]{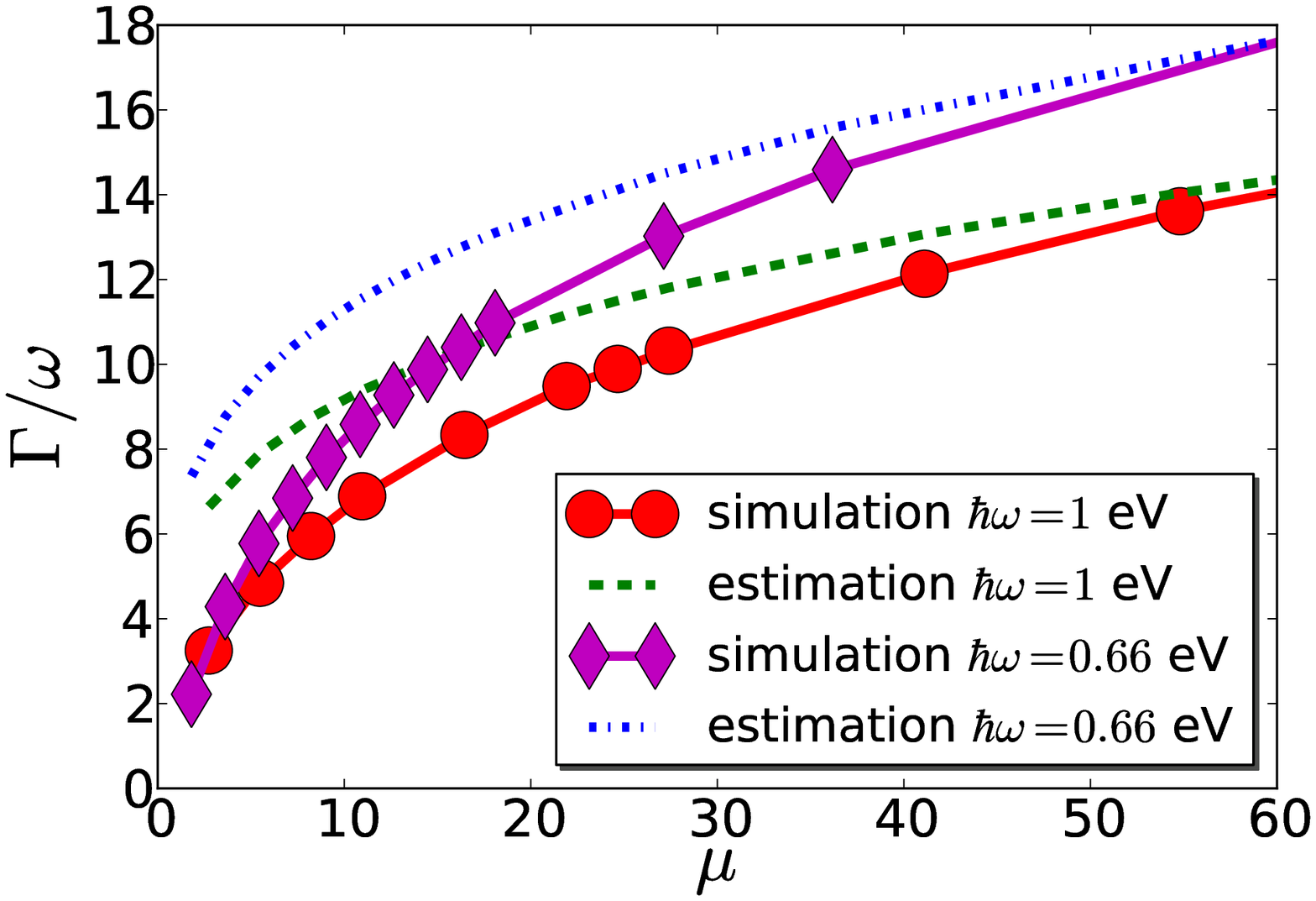}
\caption{\label{param} Left plot: Parametric studies of the mean energy $\varepsilon_e$, the mean dynamical quantum parameter $\chi_e$, and the mean angle $\theta_e$ between the momentum and the field for electrons and positrons. The ratios of the simulation results to the approximations (\ref{energies_chi_mu}), (\ref{theta_mu}) are plotted vs. the parameter $\mu$ for $\hbar\omega=1\mbox{eV}$. Right plot: Parametric study of the increment $\Gamma$ as a function of the dimensionless field strength $\mu$ for two rotation frequencies $\hbar\omega = 1\, \mbox{eV}$ and $\hbar\omega = 0.66\, \mbox{eV}$. The approximation (\ref{N(t)}) is shown by the dashed lines.}
\end{figure*}

The results of our simulations are collected in the figures ~\ref{cascade_devel}, \ref{chi_N(t)}-\ref{param}. Fig.~\ref{cascade_devel} represents a typical spatial picture of the formation and development of a cascade initiated by a positron. The electrons and positrons are deflected by the field in opposite directions, whereas the directions of propagation of photons are distributed randomly, as could be expected. For the rest of the paper we assume for all simulations in an uniformly rotating field that at $t=0$ we have a single electron at rest ($\vec{p}_e=0$) and no photons and positrons. The typical evolution of the quantum dynamical parameter $\chi_e$ of the primary electron is illustrated with the left panel of Fig.~\ref{chi_N(t)}. Before the emission of a first photon, the electron is gaining energy and its parameter $\chi_e$ is growing as the square of time in accordance with Eq.~(\ref{chi(t)1}). After the first photon emission, which for our parameters happens typically on the time scale $t_e$ smaller than $\omega^{-1}$, the curves become stochastic and consist of smooth sections with typical growth of $\chi_e$ due to acceleration by the field. These sections are separated by sudden breakdowns resulting from recoils due to successive photon emissions. Since these recoils are random, the three curves in the figure corresponding to independent simulation runs deviate at later times. After the transient period which typically lasts for several lifetimes $t_e$, the momentum losses due to quantum recoils are coming into equilibrium on the average with the trend of acceleration by the field. After that, function $\chi_e(t)$ for an individual electron describes a stationary stochastic process.

As it was predicted in Ref.~\cite{Fed}, the development of a cascade results in exponential growth with time of the total numbers of secondary hard photons and electron-positron pairs. This is illustrated with the right panel in Fig.~\ref{chi_N(t)}. The plot $N_{e^-e^+}(t)$ is given by a random stairway, with each stair corresponding to creation of a single pair. The successive stairs are well separated initially, when the total number of pairs remains small.
At later time with the number of pairs growing rapidly the stair-like structure of the lines in the plot becomes invisible and straight lines are obtained. Although these straight lines for independent simulation runs are typically different, mostly because emission of the first photon starts randomly from one simulation run to another, nevertheless their \emph{gradients} are varying weakly in different runs and can be used to determine the growth rate $\Gamma$ in Eq.~(\ref{N(t)}). For example, the growth rates extracted from the curves 1-3 at Fig.~\ref{chi_N(t)} are $\Gamma = 4.62,\,4.84$ and $4.90$, respectively.

We have studied the averages of the quantities $\chi_e$, $\varepsilon_e$ and $\theta$ over the cascade. For example, temporal evolution of the mean value
\begin{equation}\label{av_chi}
\langle \chi_e (t)\rangle=\frac1{N_{e^-}(t)}\sum\limits_{i=1}^{N_{e^-}(t)}\chi_{e\,i}(t),
\end{equation}
where $N_{e^-}(t)$ is the instant number of present electrons and $\chi_{e\,i}(t)$ is the instant value of the quantum dynamical parameter for the $i$-th electron, as depicted in the left plot of Fig.~\ref{av_chi_en}. One can see that at later times the random fluctuations are smoothed out and the quantity (\ref{av_chi}) stabilizes acquiring some definite constant value which is independent of the simulation run. The same behavior was observed for $\langle\varepsilon_e\rangle$ and $\langle\theta\rangle$, which are defined in a manner similar to Eq.~(\ref{av_chi}). The typical evolution of the averaged energy of all the components of the cascade is represented in the right plot of Fig.~\ref{av_chi_en}. At later times, the mean energies of electrons and positrons coincide as is expected from symmetry consideration, whereas the mean photon energy typically remains smaller. At the same time, the energy spectrum of created electrons and positrons is wider than the photon spectrum (see Fig.~\ref{spectr}). Both features are explained naturally by the fact that in our setup the hard photons ($\chi_\gamma\gtrsim 1$) are quickly converted into pairs which survive, whereas soft photons ($\chi_\gamma\lesssim 1$) are stable with respect to pair photoproduction and hence are accumulated. In the high energy region, all the spectra are likely to show exponential decay.

One of our main tasks was the investigation of the validity of estimations (\ref{estimations}) and (\ref{N(t)}) which were suggested previously in  Ref.~\cite{Fed} and are of crucial importance. In particular,  Eq.~(\ref{N(t)}) was serving as an argument for the estimation of the maximal value of the intensity attainable with focused laser fields. In order to test these estimations we have performed parametric studies of the stabilized values of the quantities $\langle\chi_e\rangle$, $\langle\varepsilon_e\rangle$, $\langle\theta\rangle$ and of the increment $\Gamma$. These results are presented in Fig.~\ref{param}. The left plot demonstrates the dependence of the ratios of the quantities obtained by simulations to their estimations (\ref{estimations}) on $\mu$ for fixed rotation frequency $\hbar\omega=1\mbox{eV}$. It is clear from the figure that for large values of $\mu$ each ratio acquires some definite limit of the order of unity. According to our results the formulas (\ref{estimations}) are valid up to some numerical coefficients of the order of unity, which vary no more than twice in the whole range $\mu>1$. The results of simulations for $\Gamma$ are compared with Eq.~(\ref{N(t)}) for two different rotation frequencies $\hbar\omega=0.66\mbox{eV}$ and $\hbar\omega=1\mbox{eV}$ on the right panel of Fig.~\ref{param}. One can see that for large values of $\mu$ the estimation (\ref{N(t)}) is justified with good accuracy even without any correction factor. For $\mu\lesssim 30$ formula (\ref{N(t)}) overestimates $\Gamma$ but not more than by half of an order of magnitude. This may be nevertheless crucial for the estimation of the total cascade yield due to its exponential dependence on $\Gamma$. For the particular value $\mu\approx 10$, which was exploited in Ref.~\cite{Fed}, formula (\ref{N(t)}) overestimates $\Gamma$ by approximately a factor of $1.5$. This, however, can be compensated in principle by simultaneous underestimation of the escape time in Ref.~\cite{Fed}.

In order to apply the results of our simulations to estimate the cascade yield by a realistic focused laser field, we assume that the appearing electrons and positrons are pushed away as a whole from the focus by the ponderomotive potential in radial direction with almost the speed of light. Assuming the Gaussian profile of the focused beam we can write $\mu(t)=\mu_0 e^{-c^2t^2/w_0^2}$, where $\mu_0$ is the value of the parameter $\mu$ at the center of the focus and $w_0$ is the focal radius. Then the total number of pairs produced by the cascade can be estimated to the
$$
\ln(N_{e^+e^-})\sim \int\limits_0^\infty \Gamma(\mu(t))\,dt=\Gamma(\mu_0)
\int\limits_0^\infty e^{-c^2t^2/4w_0^2}\,dt.
$$
The remaining integral defines the effective time of escape from the focus and equals $\sqrt{\pi}(w_0/c)$, i.e. is $\sqrt{\pi}\approx 1.77$ times larger than it was assumed in Ref.~\cite{Fed}. This correction almost totally cancels the overestimation of $\Gamma$ by formula (\ref{N(t)}) at $\mu\approx 10$ that we have observed in our simulations. Thus, we hope that the quantitative predictions in Ref.~\cite{Fed} must remain unaffected by our corrections.

One can see that we are currently neglecting the complicating details in our code such as, e.g. the elastic collisions, the Compton scattering, and the annihilation processes. Such phenomena must become important only at later times, when the plasma is dense enough. Though successive collisions and annihilations of the electron and positron from the same created pair may be important \cite{Kuchiev}, we are currently ignoring these effects for simplicity \footnote{Note in addition that the arguments of the paper \cite{Kuchiev} may be irrelevant to our domain of parameters, because the time of recollisions essentially exceeds the time $\sim t_e$ of hard photons emission which destroys the spatial coherence of field-driven oscillations of electron and positron.}. We ignore the higher-order processes, such as two-photon creation and the trident processes as well (see the remark at the end of Sec.~\ref{Sec:2}). All these assumptions are natural and commonly accepted in the present cascade theory \cite{Akhiez}, even though they may be revised in future studies.

Due to limitations of computer power, we currently stop our simulations after the creation of around $N_{e^-e^+}\lesssim 10^4$ pairs. This was shown to be enough to estimate the growth rate $\Gamma$, as well as to average the characteristics of a cascade over the ensemble of pairs with reasonable accuracy. In the time interval of simulation these pairs occupy a volume of the order $d^3$, where $d\sim c/\omega\sim 1\mu\mbox{m}$. This corresponds to the pair density $n_{e^+e^-}\sim 10^{16}\mbox{cm}^{-3}$. Typical values of the $\gamma$-factor for electrons and positrons are $\gamma\sim 10^3\div 10^4$, see Fig.~\ref{spectr}. This corresponds to their energies $\varepsilon_e=\gamma mc^2\sim 0.5\div 5\mbox{GeV}$. Assuming the temperature $T\sim\varepsilon_e/k\sim 10^{13}\mbox{K}$, we can estimate the Debye screening radius $r_D\sim \sqrt{kT/e^2n_{e^+e^-}}\sim 1\mbox{cm}\gg d$. The relativistic plasma frequency $\Omega_{pe}\sim c/r_D\sim 10^{10}\mbox{sec}^{-1}$ remains about five orders of magnitude smaller than the optical frequency. For these reasons we have completely neglected Coulomb interaction between electrons and positrons and all the accompanying collective plasma effects in the present simulations. However, the density of pairs is growing exponentially, $n_{e^+e^-}(t)\propto e^{\Gamma t}$, and hence $r_D\propto e^{-\Gamma t/2}$ and $\Omega_{pe}\propto e^{\Gamma t/2}$. After a relatively short period of time $\lesssim 2\pi/\omega$, when the number of pairs becomes macroscopic ($\sim 10^{11}$), the quantities $r_D$ and $\Omega_{pe}$ attain the values $d$ and $\omega$, respectively, and the collective plasma effects may come into play. Within our approximation of a homogeneous field, the total number of created pairs would be restricted by the screening of the external field by the self-field of arising plasma.

Let us note, that despite some doubts in the literature \cite{Bul}, the radiation friction is taken into account properly in our version of the algorithm by the recoils happening at the times of photon emission (see, e.g., Ref.~\cite{DiPiazza} and the Appendix \ref{sec:cascade_eqs} in our paper). Hence, there is no need to include an additional radiation friction force in the equations of motion (\ref{Newton}) for electrons and positrons, otherwise this would cause double counting. Moreover, our approach transfers the concept of classical radiation friction into the quantum domain in a correct fashion. It can be asked how the classical continuously acting radiation friction can be recovered from the sudden jumps of momentum similar to those in Fig.~\ref{chi_N(t)}. In fact this happens on the average with respect to the ensemble of Monte-Carlo realizations, since the moments of successive photon emissions are distributed randomly. At later times, when the number of created pairs becomes large, the cascade forms a representative ensemble itself and there is in principle no need for taking the average over independent realizations.

\section{Summary and Discussion}
\label{Sec:Discuss}

In this paper we have presented the first results of numerical simulations of the formation and development of electron-positron-photon cascades by initially slow electrons in a uniformly rotating homogeneous electric field. In such a situation the cascades reveal a principally new feature, i.e. the restoration of the energy and the dynamical quantum parameter due to the acceleration of electrons and positrons by the field. This feature may be of crucial importance for the whole physics of laser-matter interactions in the strong field domain, as it was demonstrated in Ref.~\cite{Fed}. We have explicitly identified this restoration mechanism in the course of our simulations. Also, our simulations clearly confirm the qualitative analysis of Ref.~\cite{Fed}, including the basic scalings (\ref{estimations}) and the estimation (\ref{N(t)}) for the cascade yield. So, they can be used to fix the remaining numerical correction factors in Eqs.~(\ref{estimations}) and (\ref{N(t)}), which turn out to be of the order of unity.

The numerical approach that we adopt is based on Monte-Carlo simulations of the cascade equations. We have shown explicitly that contrary to some recent doubts in literature \cite{Bul} such an approach incorporates radiation friction acting on individual electrons and positrons and, moreover, is doing this in a manner which is consistent with intense field QED.

The codes designed for our task can be readily adopted for simulating cascades in the laser fields with more realistic configurations, such as tightly focused Gaussian beams and pulses. This is required in order to make more definite predictions on the impact of cascade production for possible future experiments, as well as for further corrections of the maximal value of intensity that can be attained with optical lasers \cite{Fed}. Simulation of cascades in a focused laser field will be presented in a separate publication. However, let us make several brief comments about cascades in focused laser fields, possible experimental scenarios, and some yet unresolved technical problems that may require further studies.

The restoration mechanism arises due to the curvature of the trajectories of the charged particles across the field and may be sensitive to its polarization. Although we expect that this mechanism must work for generic field configurations (e.g., for generic tightly focused laser fields), there may exist several particular configurations for which the restoration mechanism does not work. For example, in an arbitrary constant electromagnetic field or a circularly polarized propagating plane electromagnetic wave the dynamical quantum parameter $\chi_e$ is conserved exactly in the course of motion. In the case of a generic propagating plane wave the amplitude of oscillations of the parameter $\chi_e$ for an initially slow electron does not exceed $E_0/E_S$, i.e., always remains smaller than unity. Another example is a linearly polarized oscillating electric field \cite{Bul}, since in this case the initially slow particles are accelerated strictly along the field and hence the growth of the transverse component of the field is absent. In some intermediate cases, e.g. for elliptical polarization or a weakly focused Gaussian beam, restoration of $\chi_e$ must exist but may be less effective than in the case of circular polarization. However, in all the cases at least the usual cascades would be caused by external high-energy electrons or hard photons passing through the high-field region transverse to the field. In this case, the cascade yield remains microscopic and would be determined by both the initial energy of an external energetic particle and the laser field strength.

In order to initiate a cascade in a tightly focused laser field, it is required to inject a primary particle into the center of a focal region. This task may be not trivial because the focal region is surrounded by a ponderomotive potential wall of the characteristic height $U_0\sim mc^2\sqrt{1+a_0^2}\approx mc^2 a_0$. The external high-energy electrons will be most likely deflected rather than penetrate inside. The most direct and elegant scenario is based on the exploration of pairs that are created spontaneously from vacuum by the laser field itself \cite{Fed}, since they are appearing exactly at the center of the focus as required. However, this possibility implies high intensities $\gtrsim 10^{26}\mbox{W/cm}^2$. Another possible resolution would be the initiation of cascades by energetic $\gamma$-quanta. In our opinion, the final conclusion whether or not cascades \emph{with macroscopic yield} can arise in \emph{generic} real experiments exploring laser-matter interaction at intensities lower than $\sim 10^{26}\mbox{W/cm}^2$ requires further studies. We note that for cascades that arise in the course of the interaction of high-intensity laser radiation with material targets it may be necessary to take the impact of ordinary cascades in matter into account as well \cite{Liang}.

If the cascade yield attains macroscopic values ($N_{e^-e+}\sim 10^{11}$), the self-field of the electron-positron plasma becomes comparable to the guiding field. In this regime screening of the external field and its absorption by the electron-positron plasma self-field will restrict further pair production. Such a regime can be simulated by combining our codes with the Particle-in-Cell (PIC) method \cite{PIC}. We hope to address this problem in one of our next publications.

\begin{acknowledgments}
We are grateful to G. Korn for his permanent interest in this work and useful advise. We are also grateful to S.V. Bulanov, J. Kirk, A. Bell and I. Sokolov for fruitful disputes that gave us several useful ideas and to E. Vilenius and S. Rykovanov for reading this manuscript and suggesting a number of stylistic corrections. This work was supported by the grant DFG RU 633/1-1, the Cluster-of-Excellence 'Munich-Centre for Advanced Photonics' (MAP), the Dynasty foundation, Russian Fund for Basic Research, the Ministry of Science and Education of the Russian Federation and the Russian Federal Program ``Scientific and scientific-pedagogical personnel of innovative Russia''.
\end{acknowledgments}

\appendix

\section{Cascade equations and radiation reaction}
\label{sec:cascade_eqs}

The cascade equations for a uniformly rotating homogeneous electric field
\begin{widetext}
\begin{eqnarray}
\frac{\partial f_{\pm}(\vec{p}_e,t)}{\partial t}\pm e\vec{E}(t)\cdot\frac{\partial f_{\pm}(\vec{p}_e,t)}{\partial \vec{p}_e}
=\int\limits w_{rad}(\vec{p}_e+\vec{p}_\gamma\to \vec{p}_\gamma)f_{\pm}(\vec{p}_e+\vec{p}_\gamma,t)\,d^3p_\gamma
-f_{\pm}(\vec{p}_e,t)\int\limits  w_{rad}(\vec{p}_e\to\vec{p}_\gamma)\,d^3p_\gamma
\nonumber\\
+\int\limits  w_{cr}(\vec{p}_\gamma\to\vec{p}_e)f_{\gamma}(\vec{p}_\gamma,t)\,d^3p_\gamma,
\label{casc1}\\
\frac{\partial f_{\gamma}(\vec{p}_\gamma,t)}{\partial t}
=\int\limits w_{rad}(\vec{p}_e\to \vec{p}_\gamma)[f_{+}(\vec{p}_e,t)+f_{-}(\vec{p}_e,t)]\,d^3p_e
-f_{\gamma}(\vec{p}_\gamma,t)\int\limits w_{cr}(\vec{p}_\gamma\to\vec{p}_e)\,d^3p_e.
\label{casc2}
\end{eqnarray}
\end{widetext}
differ from the standard equations of EAS \cite{recent,math} only by the addition of the second term to the LHS of Eq.~(\ref{casc1}), which takes electron and positron acceleration into account. Here, $f_\pm$ and $f_\gamma$ are the distribution functions for positrons, electrons and photons, respectively. In our approximation of photon and pair emission in strictly forward direction the differential probability rates are of the form
\begin{eqnarray}
w_{rad}(\vec{p}_e\to \vec{p}_\gamma)=\int\limits_0^1 d\lambda\left.\frac{dW_{rad}}{d\varepsilon_{\gamma}}
\right\vert_{\varepsilon_\gamma=\lambda\varepsilon_e}
\delta(\vec{p}_\gamma-\lambda\vec{p}_e),\nonumber\\
w_{cr}(\vec{p}_\gamma\to \vec{p}_e)=\int\limits_0^1 d\lambda\left.\frac{dW_{cr}}{d\varepsilon_e}
\right\vert_{\varepsilon_e=\lambda\varepsilon_\gamma}
\delta(\vec{p}_e-\lambda\vec{p}_\gamma),\label{w}
\end{eqnarray}
so that the integrals standing on the RHS of Eqs.~(\ref{casc1}), (\ref{casc2}) are essentially one-fold. Nevertheless, in the main case of interest, the direction of the field $\vec{E}(t)$ varies in time, so that the problem does not reduce to 1D. Note that the scalings (\ref{estimations}) can be obtained from (\ref{casc1}) and (\ref{casc2}) in the limit of large $\chi$ via dimensional analysis.

It is interesting to note that in the view of the approximation (\ref{w}) it follows from the Eqs.~(\ref{casc1}), (\ref{casc2}) that
\begin{eqnarray}
\frac{d}{dt}\left\{\int\limits \vec{p}_e(f_++f_-)
d^3p_e+\int\limits \vec{p}_\gamma f_\gamma
d^3p_\gamma\right\}\nonumber\\=e\vec{E}(t)\int\limits  (f_+-f_-)d^3p_e,\label{momentum}\\
\frac{d}{dt}\left\{\int\limits \varepsilon_e(f_++f_-)
d^3p_e+\int\limits \varepsilon_\gamma f_\gamma
d^3p_\gamma\right\}\nonumber\\=e\vec{E}(t)\int\limits  \frac{\vec{p}_e}{\varepsilon_e}(f_+-f_-)d^3p_e.\label{energy}
\end{eqnarray}
These identities ensure that in our approximation the momentum and the energy are extracted from the field only during the acceleration of electrons and positrons, i.e., both energy and momentum of the electron-positron-photon plasma are conserved during photon emission and photons to pairs conversion.

The first two terms on the RHS of the Eq.~(\ref{casc1}) describe the influence of photon emission on electrons and positrons. Let us now demonstrate that  classical radiation reaction is taken into account properly by these terms.

For this aim let us assume in what follows that the electrons are slow in the sense that distributions $f_\pm$ are restricted to such momenta for which $\chi_{e\pm}\ll 1$ (though we continue to assume that they are ultrarelativistic). In this case their motion can be described in completely classical terms. Accordingly, let us skip the third term on the RHS of the Eq.~(\ref{casc1}), which is responsible for pair production. Once this is done the total numbers of positrons and electrons $N_\pm=\int f_\pm d^3p_e$ are conserved.

The relation between the variables $\chi_\gamma$ and $x$ in Eq.~(\ref{phot_emiss}) can be expressed in the alternative form
\begin{equation}\label{kappa_x}
\chi_\gamma=\frac{x^{3/2}\chi_e^2}{1+x^{3/2}\chi_e}.
\end{equation}
It is clear from this formula by taking into account that the spectrum (\ref{phot_emiss}) of emitted photons is effectively concentrated in the range $x\lesssim 1$, that
\begin{equation}\label{kappa_x_cl}
\chi_\gamma\approx x^{3/2}\chi_e^2\lesssim\chi_e^2\ll\chi_e.
\end{equation}
As a consequence, $\vec{p}_\gamma\ll \vec{p}_e$ in the remaining integrals on the RHS of Eq.~(\ref{casc1}). Thus, the expansion
\begin{eqnarray}
w_{rad}(\vec{p}_e+\vec{p}_\gamma\to \vec{p}_\gamma)f_{\pm}(\vec{p}_e+\vec{p}_\gamma)
-w_{rad}(\vec{p}_e\to\vec{p}_\gamma)f_{\pm}(\vec{p}_e)\nonumber\\
\approx \vec{p}_\gamma\cdot\frac{\partial}{\partial\vec{p}_e}
[w_{rad}(\vec{p}_e\to\vec{p}_\gamma)f_{\pm}(\vec{p}_e)],\nonumber\\\label{expans}
\end{eqnarray}
is valid.

Consider the average momenta $\vec{P}_\pm(t)=(1/N_\pm)\int \vec{p}_e f_\pm(\vec{p}_e,t)\,d^3p_e$. By multiplying Eq.~(\ref{casc1}), truncated as described above, by $\vec{p}_e$ and integrating it over $\vec{p}_e$ by parts we have
\begin{equation}\label{eq_mot_rad}
\dot{\vec{P}}_\pm(t)=\pm e\vec{E}(t)+\langle \vec{R}\rangle_\pm (t),
\end{equation}
where
\begin{equation}\label{R}
\vec{R}(\vec{p_e})=-\int\limits \vec{p}_\gamma w_{rad}(\vec{p}_e\to\vec{p}_\gamma)\,d^3p_\gamma
\end{equation}
is the mean rate of momentum losses of electrons and positrons due to photon emission, and
$$
\langle \vec{R}\rangle_\pm (t)=\frac1{N_\pm}\int\limits \vec{R}(\vec{p_e})
f_\pm(\vec{p_e},t)
$$
are its averages over the momentum distributions of positrons and electrons, respectively.

Taking into account that $\vec{p}_\gamma\approx x^{3/2}\chi_e\vec{p}_e$ and $\varepsilon_\gamma=(\varepsilon_e/\chi_e)\chi_\gamma$ and using Eqs.~(\ref{w}) and (\ref{phot_emiss}), we are coming to
\begin{equation}\label{R1}
\vec{R}(\vec{p_e})=-\varepsilon_e\vec{p}_e\int\limits_0^{\chi_e} x^{3/2} \frac{dW_{rad}}{d\varepsilon_{\gamma}}\,d\chi_\gamma.
\end{equation}
From this point, let us pass to the integration over the variable $x$. In the approximation $\chi_e\ll 1$ we have $\chi_\gamma=x^{3/2}\chi_e^2$. Contribution to the integral in Eq.~(\ref{R1}) comes from the range $x\sim 1$. We can neglect the term $\chi_\gamma\sqrt{x}$ in the brackets in the expression (\ref{phot_emiss}) and, in addition, replace the upper limit of integration over $x$ by infinity. After these manipulations, we have
\begin{eqnarray}
\vec{R}=\frac32\frac{\alpha m^2 c^4\chi_e^2}{\hbar}\frac{\vec{p}_e}{\varepsilon_e} J,\nonumber\\
J=\int\limits_0^\infty \left\{x^2\int\limits_x^\infty {\rm Ai}(\xi)\,d\xi+2x {\rm Ai}'(x)\right\}\,dx.\label{R2}
\end{eqnarray}
The remaining integral $J$ by integration by parts and exploiting the Airy equation ${\rm Ai}''(x)-x{\rm Ai}(x)=0$ is reduced to
$$
J=-\frac23\int\limits_0^\infty x^3 {\rm Ai}(x)\,dx=-\frac49.
$$
Thus, in a view of Eq.~(\ref{chi}), we finally have
\begin{equation}\label{R3}
\vec{R}=-\frac23\frac{\alpha m^2 c^4\chi_e^2}{\hbar}\frac{\vec{p}_e}{\varepsilon_e}=
-\frac23\frac{e^4F_\perp^2\gamma_e^2}{m^2c^4} \frac{c\vec{p}_e}{\varepsilon_e}.
\end{equation}
This is exactly the main contribution to the Landau-Lifshitz (LL) force \cite{Landau2} for ultrarelativistic electrons. Other contributions to
LL equation do not appear in our derivation only because we are essentially using the constant crossed field approximation.

In the quantum case $\chi_e\gg 1$ the expansion (\ref{expans}) is not valid. Thus, radiation friction in the quantum regime cannot be described  by the concept of classical force in principle, as it was attempted to do e.g. in \cite{BellKirk1}. In addition to the advection term in the transport equation, which could be ascribed to the radiation reaction force as above, spreading of the distribution functions in momentum space becomes important as well. This spreading is associated with quantum fluctuations and is observable, e.g., as quantum excitation of synchrotron and betatron oscillations \cite{syncrotr}.


\end{document}